\documentclass{article}

\usepackage{arxiv}
\usepackage[superscript,biblabel]{cite}

\usepackage[utf8]{inputenc} 
\usepackage[T1]{fontenc}    
\usepackage{hyperref}       
\usepackage{url}            
\usepackage{booktabs}       
\usepackage{amsfonts}       
\usepackage{nicefrac}       
\usepackage{microtype}      
\usepackage{lipsum}

\usepackage{graphicx}
\usepackage{dcolumn}
\usepackage{bm}

\usepackage{amsmath}

\usepackage{mathrsfs}
\usepackage{tikz}
\usepackage{amssymb}

\usepackage{graphicx}

\usepackage[super]{natbib}

\usepackage[algoruled,vlined]{algorithm2e}

\newcommand{\DDg}[4]{{#1}_{#2}^{#3}(#4)}

\title{Investigation of Patient-sharing Networks Using a Bayesian Network Model Selection Approach for Congruence Class Models}

\author{
  Ravi Goyal\\
  Mathematica \\
  600 Alexander Park, Suite 100\\
  Princeton, NJ 08540\\
  \texttt{rgoyal@mathematica-mpr.com} \\
   \And
  Victor De Gruttola \\
  Department of Biostatistics\\
  Harvard T.H. Chan School of Public Health\\
  655 Huntington Avenue \\
  Boston, MA 02115\\
  \texttt{degrut@hsph.harvard.edu} \\
}

\begin{document}
\maketitle

\begin{abstract}
A Bayesian approach to conduct network model selection is presented for a general class of network models referred to as the congruence class models (CCMs). CCMs form a broad class that includes as special cases several common network models, such as the Erd\H{o}s-R\'{e}nyi-Gilbert model, stochastic block model and many exponential random graph models. Due to the range of models able to be specified as a CCM, investigators are better able to select a model consistent with generative mechanisms associated with the observed network compared to current approaches.  In addition, the approach allows for incorporation of prior information.  We utilize the proposed Bayesian network model selection approach for CCMs to investigate several mechanisms that may be responsible for the structure of patient-sharing networks, which are associated with the cost and quality of medical care. We found evidence in support of heterogeneity in sociality but not selective mixing by provider type nor degree.
\end{abstract}


\section{Introduction}

There is a growing body of research that leverages administrative claims data to identify connections among medical providers; two providers are deemed to have a connection when both treat the same patient as indicated by medical claims. Such connections have been shown to imply clinical relationships among providers.\citep{barnett2011mapping, dugoff2018geographic}   The collection of such connections has been referred to as patient-sharing networks.\citep{landon2012variation, dugoff2018scoping} Investigating the relationship between such networks and patient health outcomes is an emerging area of research; see DuGoff et al. (2018) for a systematic review of the literature.\citep{dugoff2018scoping} The most common theoretical explanation for this association is that the networks represent aspects of collaboration, continuity, and care  coordination;\citep{barnett2011mapping, pollack2015patient} these aspects may be especially important for patients with multiple chronic conditions, who account for a high percentage of health care costs.\citep{bodenheimer2009care} Health outcomes that have been studied in this regard include cost,\citep{barnett2012physician, landon2018patient, uddin2013mapping, uddin2016exploring, uddin2013study, uddin2011effect, uddin2015impact, hussain2015collaboration, pollack2013patient, agha2018team} utilization,\citep{pollack2013patient, pollack2014s} patient-reported outcomes,\citep{carson2016characterizing, carson2016outcome} quality of care,\citep{dugoff2018geographic, hollingsworth2016association} and mortality.\citep{hussain2015collaboration, hollingsworth2016association} Even as such evidence grows, there remain gaps regarding how to make use of knowledge about the patient-sharing networks to improve health care. In particular, neither how patient-sharing networks evolve in response to different incentives nor how to use them to develop or evaluate interventions is well understood.\citep{dugoff2018scoping} Addressing these gaps requires knowledge of the generative mechanisms governing the evolution of the patient-sharing network. 

In the manuscript, we present an approach that facilitates evaluation of the evidence supporting hypotheses that observed networks were generated by specified  mechanisms. We apply this approach to evaluate the presence of heterogeneity in sociality and selective mixing in patient-sharing networks; details about these mechanisms are provided in Section 3. The approach requires 3 steps: 1) development of network models consistent with potential generative mechanisms, 2) evaluation of the evidence supporting the proposition that a given network model generated the observed network, and 3) selection of the model with the greatest evidence as measured by their posterior probabilities. The posterior probability estimation is based on a novel approach to Bayesian model selection among a class of models--denoted as congruence class models (CCMs) for networks.\citep{goyal2014sampling} CCMs form a broad class that includes as special cases several common network models, such as the Erd\H{o}s-R\'{e}nyi-Gilbert (ER) model and stochastic block (SB) model as well as many exponential random graph models (ERGMs)--one of the most commonly used network models in social network analysis. See Goldenberg et al. for a review of these models.\citep{goldenberg2010survey} 

The ability of CCMs to generalize such a broad set of network models arises from the flexibility in specifying the probability distribution associated with network properties included in a model. This flexibility enables investigators to develop network models that more closely correspond to potential network generation mechanisms than do current classes of network models. In this regard, it allows investigators to make better use of the totality of their knowledge of the generative mechanism. 

Bayesian network model selection has shown to be challenging because of the computational burden of estimating the likelihood--and therefore the posterior probability--for each candidate network model.\citep{caimo2013bayesian} The likelihood of the CCM requires the computation of the number of graphs associated with the observed values of network properties, this problem is referred to as graph enumeration. In this manuscript, we make use of a recently developed general recursive formula to estimate the number of labeled graphs for given values of graph properties that makes the computational burden feasible for mid-size networks, i.e. several thousand nodes.\cite{goyalgraphenumeration}

The next section provides background details on CCMs and their relationship to other network models (section 2.1), Bayesian model selection (section 2.2), the general graph enumeration recursive formula (section 2.3), and relevant research on network model selection (section 2.4). Section 3 introduces patient-sharing networks and a description of several competing network models representing distinct generative mechanisms. The evaluation of the models and results of the model selection approach is presented in Section 4. The paper concludes with a discussion and further research. 

\section{Background}

\subsection{Congruence Class Models for Networks}

We denote a network as $g = (V_g, E_g)$, where $V_g$ and $E_g$ are the vertex and edge sets of $g$. Let $n$ represent the number of vertices in $g$ and $\vert Z \vert$ denote the size of set $Z$; therefore, $n = \vert V_g \vert$. We represent a network $g$ as an adjacency matrix. Let $g[i,j] = 1$ indicate that there is an edge between $i$ and $j$, where $i, j \in \{v_1,\cdots, v_{n}\}$, whereas $g[i,j] = 0$ indicates that there is no edge. 

Let $\mathscr{G}_n$ denote the space of all potential networks with $n$ vertices. Let $\phi$ denote an algebraic map from $\mathscr{G}_n$ to network summary statistics (e.g., degree distribution or degree mixing) and let $c_{\phi}(x) = \{g : \phi(g) = x, g \in \mathscr{G}_n\}$ denote the inverse image associated with $\phi$. We refer to these inverse images of singleton sets as congruence classes;\citep{goyal2014sampling} they also have been referred to as fibers in algebraic statistics literature.\cite{petrovic2017survey} We use $\phi$ to represent the mapping as well as the associated network property being calculated. Let $\vert c_{\phi}(x) \vert$ denote the number of graphs for which network property $\phi$ equals $x$; this quantity has been referred to as a volume factor.\citep{Shalizi13}

The probability distribution on $\mathscr{G}_n$ for the CCM is specified by the probability mass function (PMF) on the congruence classes defined by $\phi$; we denote this PMF as $P_{\phi}$. $P_{\phi}(x \vert \theta)$ denotes the total probability of all networks that are elements in $c_{\phi}(x)$ given $\theta$, i.e.,

\begin{equation} \label{eq:networkprob_static_pc}
P_{\phi}(x \vert \theta) = \sum_{g \in c_{\phi}(x)} P_{\mathscr{G}_n}(g \vert \theta),
\end{equation}

\noindent where $ P_{\mathscr{G}_n}(g)$ is the probability of graph $g$. CCMs assume that two networks within a congruence class have the same probability of being observed; this assumption is also present in common network models including the ER model, SB model, and ERGM. Therefore, the probability distribution on $\mathscr{G}_n$ for a CCM is the following:

\begin{equation} \label{eq:networkprob_static}
P_{\mathscr{G}_n}(g \vert \theta) =\left(\frac{1}{\vert c_{\phi}(\phi(g)) \vert} \right) P_{\phi}(\phi(g) \vert \theta).
\end{equation}

CCMs are able to represent several common network models because of the flexibility in the specification of $P_{\phi}$. To illustrate this flexibility, we consider as an example the specification of a  probability distribution identical to that of the ER model.  To do so, we let $\phi_1$ be the mapping from a network $g$ to its number of edges, i.e., $\vert E_g \vert$,  and let $P_{\phi_1} (\phi_1(g) \vert \theta = p)$ equal the following: 

\begin{equation} \label{eq:ERG_CCM_prob}
P_{\phi_1} (\phi_1(g) \vert \theta = p) = p^{\phi_1(g)}(1-p)^{{n \choose 2} - \phi_1(g)} {{n(n-1)/2} \choose \phi_1(g)}.
\end{equation}

For ERGMs, networks in the same class have the same probability;\cite{petrovic2017survey} hence there exists a $P_{\phi}$ such that a CCM and an ERGM assign the same probability distribution on $\mathscr{G}_n$. However, CCMs provide additional flexibility in modeling $P_{\mathscr{G}_{n}}$ compared to ERGMs. ERGMs require $P_{\phi}$ to have a specific functional form, while the CCM does not place restrictions on this form.

\subsection{Bayesian Model Selection}

Bayesian model selection identifies the model with the highest posterior probability among a set of candidate models, which we denote as $\{m_1, \cdots, m_k\}$. This section derives the posterior probabilities for a set of potential CCMs.

Let $I_{m_h}$ be an indicator variable that $m_h$ is the correct model for the observed network. Equation~\ref{eq:model_post} shows the posterior probability for model $m_h$ given an observed network $g$:

\begin{equation} \label{eq:model_post}
P(I_{m_h}=1 \vert g) = \frac{p(g \vert I_{m_h}=1) p(I_{m_h} = 1)}{\sum_{j=1}^k p(g \vert I_{m_j}=1) p(I_{m_j}=1)},
\end{equation}

\noindent where $p(g \vert I_{m_h}=1)$ and $p(I_{m_h}=1)$ are the model evidence (shown below) and prior probability, respectively, for model $m_h$. Equation~\ref{eq:model_evidence} shows the model evidence for model $m_h$:

\begin{equation} \label{eq:model_evidence}
P(g \vert I_{m_h}=1) = \int_{\theta_h} p(g \vert \theta_h, I_{m_h}=1) p(\theta_h \vert I_{m_h}=1) d\theta_h,
\end{equation}

\noindent where $\theta_h$ are the parameters for model $m_h$, $p(g \vert \theta_h, I_{m_h}=1)$ is the likelihood, and $p(\theta_h \vert I_{m_h}=1)$ is the prior distribution of the parameters of model $m_h$. 

The model evidence for a CCM can be derived by substituting the CCM likelihood (Equation~\ref{eq:networkprob_static}) into the general equation for model evidence (Equation~\ref{eq:model_evidence}) as shown in Equation~\ref{eq:model_evidence_CCM_1}:

\begin{equation} \label{eq:model_evidence_CCM_1}
P(g \vert I_{m_h}=1) = \int_{\theta_h} \left(\frac{1}{\vert c_{\phi_h}(\phi_h(g)) \vert} \right) P_{\phi_h}(\phi_h(g) \vert \theta_h) * p(\theta_h \vert I_{m_h}=1) d\theta_h. 
\end{equation}

\noindent The volume factor is not dependent on $\theta_h$ and, therefore can be brought outside of the integral as shown below in Equation~\ref{eq:model_evidence_CCM}:

\begin{equation} \label{eq:model_evidence_CCM}
P(g \vert I_{m_h}=1) = \left(\frac{1}{\vert c_{\phi_h}(\phi_h(g)) \vert} \right) \int_{\theta_h} P_{\phi_h}(\phi_h(g) \vert \theta_h) * p(\theta_h \vert I_{m_h}=1) d\theta_h
\end{equation}

This integral can be computed using standard approaches;\citep{davis2007methods} however, our computation makes use of recent research to estimate the volume factor.\cite{goyalgraphenumeration} In the next section, we present a summary of this work.

\subsection{Graph Enumeration}

Equation~\ref{eq:recursive_formula} provides a recursive formula to estimate the number of graphs, $\vert c_{\phi}(x_k) \vert$, with specific value(s), $x_k$, for particular network properties, $\phi$:

\begin{equation}\label{eq:recursive_formula}
\vert c_{\phi}(x_k) \vert = r_{\phi}(x_k, x_{k-1}) * \vert c_{\phi}(x_{k-1}) \vert, 
\end{equation}

\noindent where $r_{\phi}(x_k, x_{k-1})$ is the ratio between the sizes of congruence class $c_{\phi}(x_{k})$ and $c_{\phi}(x_{k-1})$, i.e.,

\begin{equation}\label{eq:recursive_formula_ratio}
r_{\phi}(x_k, x_{k-1}) = \frac{\vert c_{\phi}(x_{k}) \vert }{\vert c_{\phi}(x_{k-1}) \vert}.
\end{equation}

Methods for evaluation of the recursive formula, Equation~\ref{eq:recursive_formula_ratio}, have been developed for a range of network properties of interest to social network analysis, including number of edges, mixing by nodal covariates, degree distribution, and degree mixing.\cite{goyalgraphenumeration} 

\subsection{Previous Research on Network Model Selection}

Compared to other areas of network analysis, there has been relatively little published research on model selection. For ERGMs, Caimo and Friel (2013) developed a Bayesian model selection method based on an extension of their reversible jump Markov chain Monte Carlo algorithm that estimates the posterior probability for each model.\citep{Caimo11, caimo2013bayesian} Thiemichen et al. (2016) presented a method that applies a Laplace approximation to estimate the Bayes factor, which allows for model selection in the Bayesian paradigm.\citep{thiemichen2016bayesian} However, these approaches to model selection for ERGMs suffer from a high complexity cost. In order to reduce the computational burden, Bouranis et al. (2017) and Bouranis et al. (2018) proposed alternative approach to Bayesian inference for ERGMs based on adjusting the pseudo-posterior distribution or pseudo-likelihood, respectively.\citep{bouranis2017efficient, bouranis2018bayesian} However, they note that procedures for approximating a solution to the likelihood equation are more challenging for larger datasets.\citep{bouranis2018bayesian}

Even if the computational issues of Bayesian model selection for ERGMs are overcome, the ERGMS still have severe limitations in the types of generative mechanisms that they can model. The functional form of ERGMs only allows for investigators to specify the mean of the probability distribution for each network property. Such restriction results in the inability of ERGMs to model some common generative mechanisms.\citep{goyalMPMC}

Yan et al. (2014) presents a frequentist approach to select between the stochastic block model and degree-corrected block model using a likelihood ratio test.\citep{yan2014model} Their approach is limited to comparing these two nested models. Nonetheless, they do consider a setting with additional complexity in that they assume that the mapping from vertices to blocks is unknown and must be estimated. There has also been research in a closely related area of assessing model fit for networks.\citep{hunter2008goodness, schweinberger2012statistical, gross2017goodness}

\section{Patient-Sharing Networks}

For each US state, we analyze the patient-sharing network in order to investigate the generative mechanisms underlying this network. State networks include all resident providers who share Medicare patients. The next section provides a description of the data used to generate the patient-sharing network for each state. The subsequent section provides details of generative mechanisms we investigate and their associated CCM.

\subsection{Data}

Our analysis uses two publicly available data sets from the Center for Medicare and Medicaid Services (CMS).\citep{CMS2015referral, CMS2015med} The first identifies edges between providers; we define two providers as connected if a Medicare patient encounters both within a 30-day interval. CMS provides variations of this data sets for which the time interval between encounters can be set at 30-, 60-, 90-, or 180-days. We use the 30-day interval, as it is the most conservative for implying that two providers are coordinating care for a patient; this choice is intended to reduce the number of ``spurious'' edges that arise from providers treating distinct aliments and do not need to coordinate care.\citep{barnett2012physician, an2018analysis} Our analysis assumes that providers are connected if they share at least one patient; previous studies have used this threshold as well as other thresholds.\cite{moen2016analysis, lee2011social} Our analysis investigates networks that arose in the first three quarters of 2015--the most recent publicly available data set. 

The second data set, called Medicare Provider Utilization and Payment Data for 2015, lists the geographic location and medical specialty of the provider. This information was used to filter our patient-sharing network based on the state the provider resides as well as label providers as either primary or specialty care.

\subsection{Network Mechanisms and Models}

We investigate several CCMs, denoted as $m_1$-$m_5$, that model mechanisms that may be responsible for the resultant patient-sharing network. In this section, we introduce the models, and in Section 4, present the results of selecting among the models. The first two models ($m_1$ and $m_2$) are associated with mechanisms in which edges form at random; these are presented as null models and used to investigate the importance of prior information. Model $m_3$ is associated with heterogeneity in sociality, whereas models $m_4$ and $m_5$ are associated with two different types of selective mixing.

\subsubsection{Null Model without and with Prior Information ($m_1$ and $m_2$)}

The simplest mechanism we consider assumes that each pair of providers form a connection with a fixed probability, $p$,  that is independent of all other edges; networks are generated based on this assumption.  Hence, the mechanism corresponds to the ER model, which is a commonly used as a null network model. We investigate two CCMs ($m_1$ and $m_2$) that are both based on the this mechanism, but vary in their prior information; $m_1$ represents an ER model with no prior information, whereas $m_2$ represents the same mechanism as $m_1$, except in that we include an informative prior based on patient-sharing networks from states other than the one of focus (we consider these analyses by state).  Therefore for both $m_1$ and $m_2$, $\phi_1(g) = \vert E_g \vert$ and $P_{\phi_1} (\phi_1(g) \vert \theta_h)$ is equal to Equation~\ref{eq:ERG_CCM_prob}. Assuming no prior information on $p$, the model evidence for $m_1$ is shown below in Equation~\ref{eq:model_evidence_m1}:

\begin{equation} \label{eq:model_evidence_m1}
P(g \vert I_{m_1}=1) = \left(\frac{1}{\vert c_{\phi_1}(\phi_1(g)) \vert} \right)  * \int_{p} p^{\phi_1(g)}(1-p)^{{n \choose 2} - \phi_1(g)} {{n(n-1)/2} \choose \phi_1(g)} dp. 
\end{equation}

For $m_2$, we assume that the prior information for $p$ follows a beta distribution, $Beta(\alpha_2, \beta_2)$. Therefore, the model evidence for $m_2$ is shown below in Equation~\ref{eq:model_evidence_m2}:

\begin{equation}\label{eq:model_evidence_m2}
\begin{split}
P(g \vert I_{m_2}=1) =& \left(\frac{1}{\vert c_{\phi_1}(\phi_1(g)) \vert} \right) * \int_{p} p^{\phi_1(g)}(1-p)^{{n \choose 2} - \phi_1(g)} {{n(n-1)/2} \choose \phi_1(g)}\\
& * \frac{\phi_1(g)^{\alpha_2 -1} * (1-\phi_1(g))^{\beta_2-1}}{B(\alpha_2, \beta_2)} dp.
\end{split}
\end{equation}

\subsubsection{Sociality ($m_3$)} 

Sociality is defined as the propensity for an individual to create connections;\citep{goodreau2009birds} our goal is to evaluate whether there is heterogeneity among individuals in sociality. As in developing any generative model, there are two steps: The first is identifying the important covariates to include in the model. The second is modeling the relationship between the covariates and the outcome of interest by choosing the appropriate probability distribution for this relationship. In our setting, these steps consist of identifying network properties, e.g., degree distribution, and then associating these properties with a probability distribution that is consistent with beliefs regarding the mechanism underlying the generation of the network.

In any realization (whether observed or simulated) of a network, heterogeneity in sociality would be reflected in the degree distribution. To investigate this issue in the patient-sharing network, we consider a model ($m_3$) that includes degree distribution as the sole network property; this selection corresponds to the first step. As Goodreau et al. (2009) note, sociality is not synonymous with degree; the former is a feature of the network generating mechanisms, whereas the latter is a feature of any given realization of the mechanism.\citep{goodreau2009birds}

For the second step, the choice of appropriate probability distribution associated with degree distribution depends on the specific mechanisms hypothesized to generate the network.  For example, a common feature in social systems is the concentration of influence to a few individuals through mechanisms that encourage preferential attachment--the mechanism wherein providers form connections with others based on a probability proportional to the number of connections the providers already have (i.e., degree).\citep{price1976general, BA99} This type of mechanism generates networks with degrees following a fat-tailed distribution--specifically a power-law distribution. Many real-world networks can be modeled using this mechanism, which leads to a network wherein many nodes have a moderate number of edges and fewer nodes have a large number. Other mechanisms may result in different distributions. For example, mechanisms based on the non-equilibrium theory can result in an exponential distribution for the degrees.\citep{deng2011exponential} CCMs provide the flexibility that enables investigates to select the most appropriate probability distribution. 

In order to specify $m_3$, we introduce some notation. Let the degree of vertex $i$, denoted as $\DDg{d}{i}{}{g}$, be the number of edges between that vertex and others; hence, $\DDg{d}{i}{}{g} = \sum_j g[i,j]$. Let $\DDg{d}{}{}{g}= (\DDg{d}{1}{}{g},\cdots, \DDg{d}{n}{}{g})$ represent the vector of degrees of nodes in set $V_g$, commonly referred to as a degree sequence. The degree distribution, denoted as $\DDg{D}{}{}{g}$, is a vector such that the $k^{th}$ entry represents the number of vertices having degree $k$, i.e., $\DDg{D}{k}{}{g} = \sum_{i=1}^{n_t} I_{\{\DDg{d}{i}{}{g} = k\}}$. Let $\phi_3$ be the mapping from a network $g$ to its degree distribution, i.e., $\phi_3(g) = \DDg{D}{}{}{g}$.

Model $m_3$ represents networks generated such that the degrees follow an exponential distribution with scale parameter $\frac{1}{\lambda}$; that is $d_i \sim exp(\lambda)$; one could also investigate alternative models, such as where the degrees follow a power-law distribution. Again, we include prior information based on the patient-sharing networks from the states other than the one of focus. We assume the prior information for $\lambda$ follows a normal distribution. The model evidence for $m_3$ is shown below in Equation~\ref{eq:model_evidence_m3}:

\begin{equation}\label{eq:model_evidence_m3}
\begin{split}
P(g \vert I_{m_3}=1) ={}& \left(\frac{1}{\vert c_{\phi_3}\phi_3(g) \vert} \right) * \int_{\lambda} \prod_{i=1}^n \prod_{j=0}^{\DDg{D}{i}{}{g}} \lambda exp^{-\lambda * i}\\
& * \frac{1}{\sqrt{2\pi\sigma}}exp\left[-\frac{1}{2}\left(\frac{\lambda-\mu}{\sigma} \right)^2 \right] d\lambda.
\end{split}
\end{equation}

\subsubsection{Selective Mixing ($m_4$ and $m_5$)}

The resultant patient-sharing network may also be influenced by the presence of mechanisms by which providers form connections based on one or more of their individual characteristics. Often mixing is assortative, but it can also be dissasortative; the former implies preferential formation of connections between individuals with similar characteristics, and the latter, between individuals with contrasting characteristics.\citep{goodreau2009birds} Model $m_4$ investigates selective mixing by specialty, whereas model $m_5$ investigates mixing by the number of connections, i.e., degree of a provider.

For model $m_4$, we consider mixing between primary and specialty care providers. Let $\DDg{MM}{}{}{g}$ be a matrix representing the mixing pattern of network $g$, where the entry $\DDg{MM}{k,l}{}{g}$ is the total number of edges between a vertex with characteristic $k$ and vertex with characteristic $l$; in our application, we are interested in the characteristic indicating the provider type (primary or specialty care). To investigate mixing by provider type in the patient-sharing network, we consider a model ($m_4$) that includes mixing matrix as the sole network property; this selection corresponds to the first step in developing a generative model. For the second step, we have model $m_4$ include three independent variables representing the proportion of edges that link: 1) a primary care provider to another primary care primary, 2) primary care provider to a specialty provider, and 3) specialty provider to another specialty provider. Each of these variables follow a binomial distribution with parameters, $p_{4,pp}, p_{4,ps}$, and $p_{4,ss}$, respectively. We assume the prior information for these parameters, based on the patient-sharing networks from the states other than the one of focus, follow beta distributions, denoted as $Beta(\alpha_{4,pp}, \beta_{4,pp})$, $Beta(\alpha_{4,ps}, \beta_{4,ps})$, and $Beta(\alpha_{4,ss}, \beta_{4,ss})$. Therefore, the model evidence for $m_4$ is as follows:

\begin{equation}\label{ eq:model_evidence_m4}
P(g \vert I_{m_4}=1) = \left(\frac{1}{\vert c_{\phi_4(g)} \vert} \right) * \int_{p_{4,pp}} \int_{p_{4,ps}} \int_{p_{4,ss}} f_{4,pp} * f_{4,ps} * f_{4,ss} dp_{4,pp} dp_{4,ps} dp_{4,ss},
\end{equation}

\noindent where,

\begin{equation}
f_{4,kl} = p_{4,kl}^{\phi_{4,kl}(g)}(1-p_{4,kl})^{{\psi_{4,kl}(g) \choose 2} - \phi_{4,kl}(g)} {\psi_{4,kl}(g) \choose \phi_{4,kl}(g)} * \frac{\phi_{4,kl}(g)^{\alpha_{4,kl} -1} * (1-\phi_{4,kl}(g))^{\beta_{4,kl}-1}}{B(\alpha_{4,kl}, \beta_{4,kl})}; 
\end{equation}

\noindent $\phi_{4,kl}(g)$ is the entry $\DDg{DMM}{k,l}{}{g}$, that is number of edges in between providers specified by $kl$; and $\psi_{4,kl}$ is number of possible edges between providers of type $k$ and $l$.

Model $m_5$ represents that networks are generated based on the mechanism of selective mixing by degree. We evaluate the presence of this mechanism by modeling the degree mixing matrix, denoted as $\DDg{DMM}{}{}{g}$; the entry $\DDg{DMM}{k,l}{}{g}$ is the total number of edges between vertices of degrees $k$ and $l$. Let $\phi_5(g) = \DDg{DMM}{}{}{g}$. As with the previous models, we need to select a probability distribution, $P_{\phi_5}$, for the selected network property based on our hypothesized network generative mechanisms. For $m_5$, we assume that the proportion of edges between nodes of degrees $k$ and $l$ ($k \leq l$) is based on the following logistic model: 

\begin{equation}\label{eq:model_evidence_m5_1}
logit(p_{k,l}) = \beta_0 + \beta_1*k + \beta_2* l 
\end{equation}

We assume the prior information for $\beta_0, \beta_1$, and $\beta_2$ follows a multivariate normal distribution; to minimize the effects of noise, our estimation of the prior only includes degrees up to $300$. Therefore, the model evidence for $m_5$, assuming that the entries of the degree mixing matrix follow a multinomial distribution, is as follows:

\begin{equation}\label{ eq:model_evidence_m5_2}
\begin{split}
P(g \vert I_{m_5}=1) ={}& \left(\frac{1}{\vert c_{\phi_5}\phi_5(g) \vert} \right) \\
& * \int_{\beta} \frac{n!}{\prod_{k=1}^{n} \prod_{l \leq k} \DDg{DDM}{k,l}{}{g}} \prod_{k=1}^{n} \prod_{l \leq k} p_{k,l}^{\DDg{DDM}{k,l}{}{g}}\\
& * \frac{(2\pi)^{-\frac{3}{2}}}{det(\Sigma)}exp\left[-\frac{1}{2}(\beta-\mu)\Sigma^{-1}(\beta-\mu)\right] d\beta.
\end{split}
\end{equation}

\section{Investigation of Patient-Sharing Networks}

In the sections below, we present findings on the value of prior information as well as whether there is evidence of sociality and selective mixing. We investigate these questions for each of the 50 states. However, we first present our findings for the state of Wyoming--chosen because a small number of providers reside in the state and, therefore, easier to visualize compared to other states.

\subsection{Investigation of Wyoming}

\begin{figure}[t]
\centerline{\includegraphics[width=342pt]{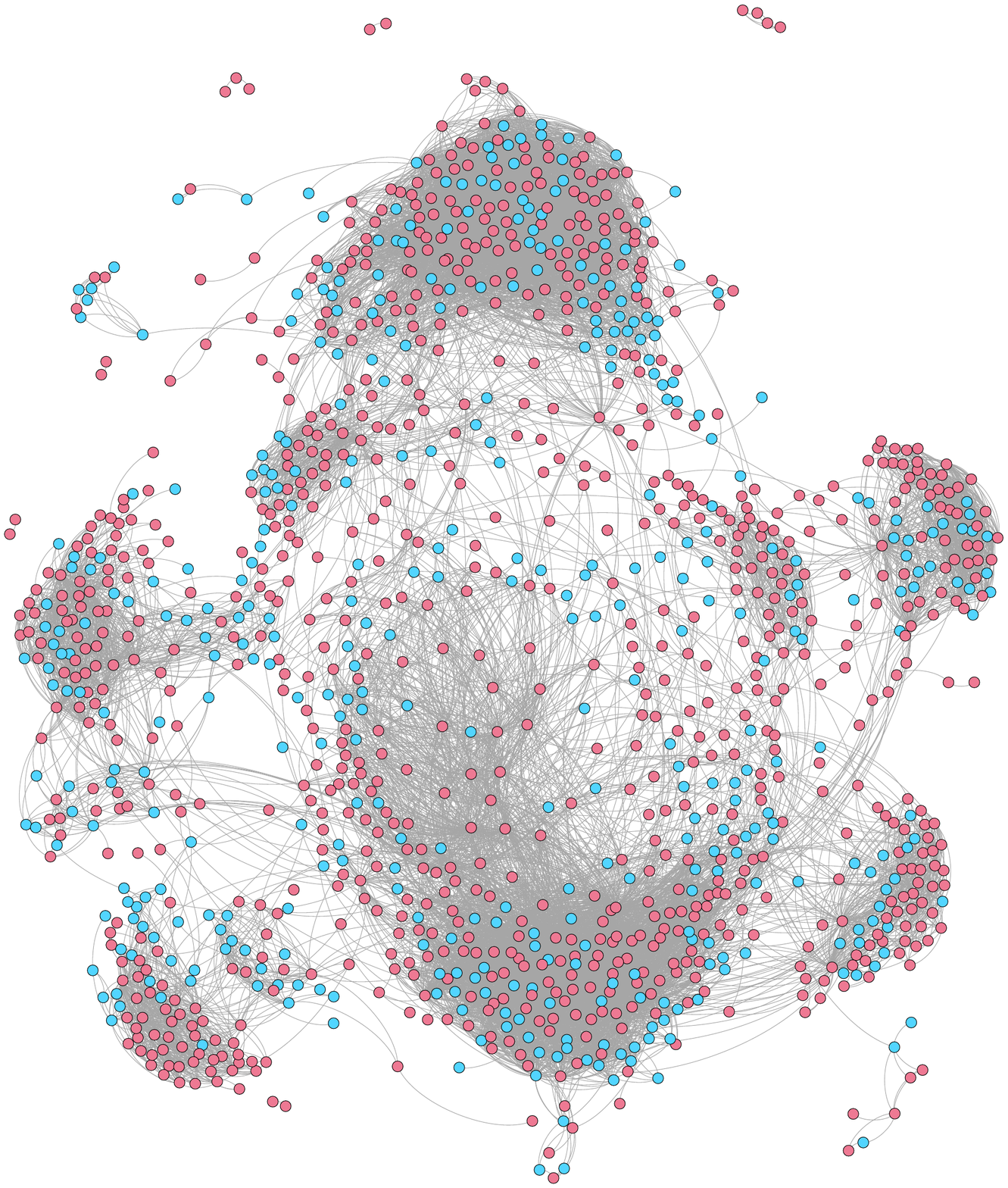}}
\caption{A visualization of the patient-sharing network for providers that share Medicare patients and reside in Wyoming. The nodes represent providers and colored based on whether they are designated as primary (blue) or specialty (red) care. The edges between the nodes represent that the providers have at least one shared patient.\label{fig:WY_patient_sharing_net}}
\end{figure}

In 2015, the state of Wyoming had 1283 medical providers that share Medicare patients; we designated 412 and 871 as primary care and as specialty providers, respectively, based on their provider type in the Medicare Provider Utilization and Payment Data. In 2015, there were 12,749 connections among these providers based on shared patients. Figure~\ref{fig:WY_patient_sharing_net} presents a visualization of the patient-sharing network for these providers. The nodes represent providers and colored based on whether they are designated as primary (blue) or specialty (red) care. The edges between the nodes represent that the providers have at least one shared patient; we denote the patient-sharing network for Wyoming as $g_{WY}$. 

\subsubsection{Value of Prior information: Comparing $m_1$ and $m_2$}

The only difference between $m_1$ and $m_2$ is the inclusion of prior information based on data from the other 49 states. By comparing such models, one can gain insight into the importance of prior information on model selection. In order to evaluate the model evidence of $m_1$ and $m_2$, we calculated the log size of congruence class $c_{\phi_1}(\phi_1(g_{WY}))$, i.e., $log \vert c_{\phi_1} (\vert E_{g_{WY}} \vert)  \vert$, as $65766.28$. The estimated model evidence for models $m_1$ and $m_2$ are $P(g_{WY} \vert I_{m_1} = 1) = 1.42\times10^{-28568}$ and $P(g_{WY} \vert I_{m_2} = 1) = 6.75\times10^{-28568}$, respectively. This results in posterior probability estimates of $P(I_{m_1} = 1 \vert g_{WY}) = 17.4\%$ and $P(I_{m_2} = 1 \vert g_{WY}) = 82.6\%$. Therefore, the inclusion of prior information alters the model evidence and poster probabilities; for the remaining comparisons, we use $m_2$.

\subsubsection{Evidence of Sociality}

\begin{figure}[t]
\centerline{\includegraphics[width=342pt]{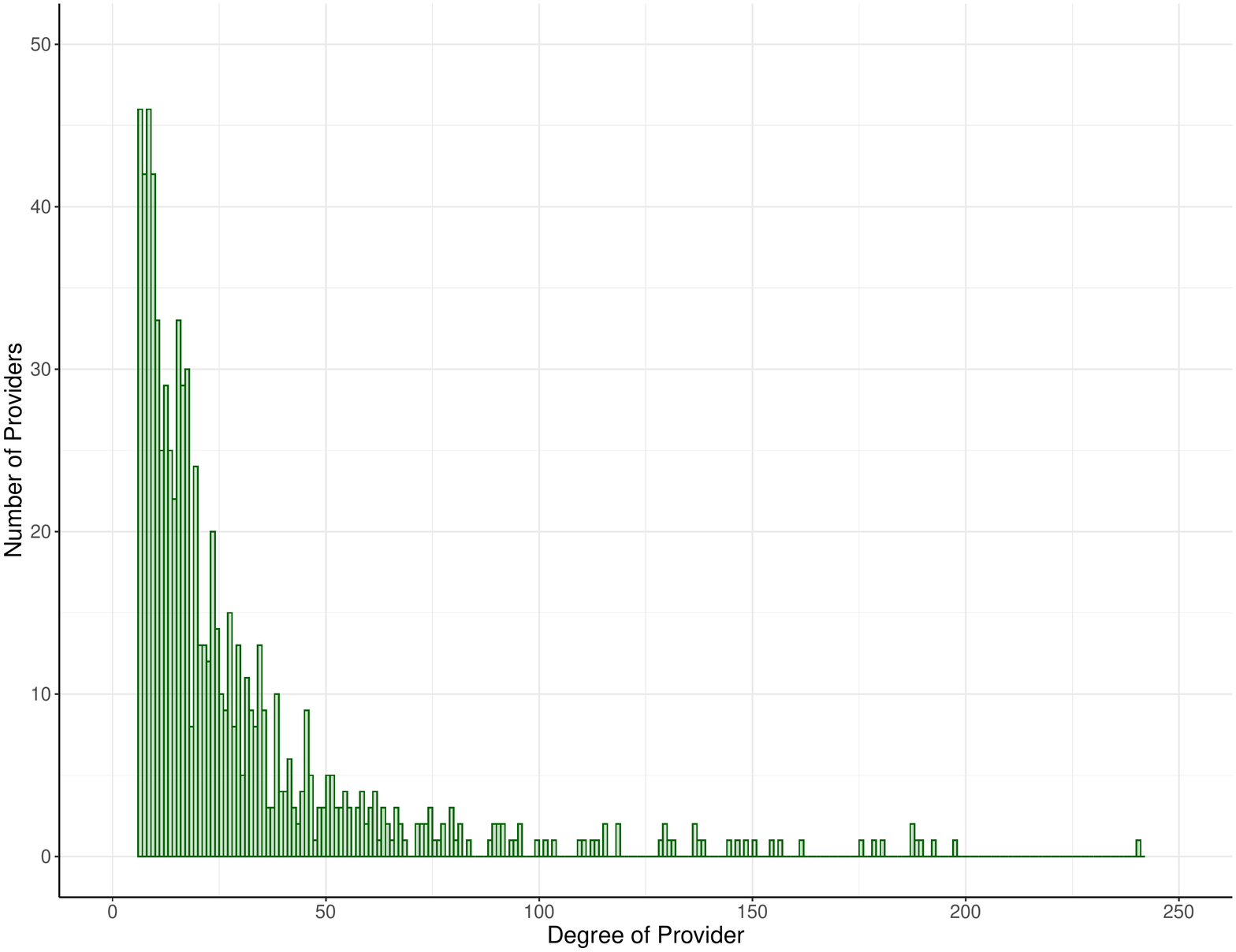}}
\caption{A visualization of the degree distribution for the patient-sharing network for providers that share Medicare patients and reside in Wyoming.\label{fig:WY_patient_sharing_deg_dist}}
\end{figure}

To investigate evidence for heterogeneity among individuals in their propensity to create connections, we assess if the observed degree distribution, shown in Figure~\ref{fig:WY_patient_sharing_deg_dist}, differs from the null model with prior information. This assessment is based on comparing models $m_2$ and $m_3$ as models $m_4$ and $m_5$ are not relevant for investigating heterogeneity of sociality. In order to evaluate the model evidence of $m_3$, we estimated the log size of congruence class $c_{\phi_3}(\phi_3(g_{WY}))$, i.e., $log \vert c_{\phi_3}(\DDg{D}{}{}{g_{WY}}) \vert$, as $49550.14$ based on the recursive algorithm described in section 2.4. The estimated model evidence for $m_3$ is $P(g_{WY} \vert I_{m_1} = 1) = 2.43\times10^{-23919}$. This results in estimated posterior probabilities of $P(I_{m_2} = 1 \vert g_{WY}) \leq 0.01\%$ and $P(I_{m_3} = 1 \vert g_{WY}) \geq 99.9\%$ when comparing only models $m_2$ and $m_3$. Therefore, there is evidence of heterogeneity in sociality contributing to the patient-sharing network in Wyoming.

\subsubsection{Evidence of Selective Mixing by Provider Type}

To investigate whether there is evidence of selective mixing by provider type, we compare models $m_2$ and $m_4$ as the other models are not relevant for this assessment.  In order to evaluate the model evidence for $m_4$, we estimated the log size of congruence class $c_{\phi_4}(\phi_4(g_{WY}))$, i.e., $log \vert c_{\phi_3}(\DDg{MM}{}{}{g_{WY}}) \vert$, as $65293.53$ based on the recursive algorithm. The estimated model evidence for $m_4$ is $P(g_{WY} \vert I_{m_4} = 1) = 1.46\times10^{-56988}$. This results in posterior probability estimates of $P(I_{m_2} = 1 \vert g_{WY}) \geq 99.9\%$ and $P(I_{m_4} = 1 \vert g_{WY}) \leq 0.01\%$. Therefore, there is little evidence for selective mixing by provider type for Wyoming. It is important to note that this conclusion does not assure that mixing by provider type is not a significant influence on the network structure; it only means that our model $m_4$--as specified by our choice of $P_{\phi_4}$--did not lead to a better model compared to our choice of null model.

\subsubsection{Evidence of Selective Mixing by Degree}

To assess evidence for selective mixing by degree, we must control for the distribution of degrees. Therefore, we compare models $m_3$ and $m_5$. In order to evaluate the model evidence of $m_5$, we estimated the log size of congruence class $c_{\phi_5}(\phi_5(g_{WY}))$, i.e., $log \vert c_{\phi_3}(\DDg{DMM}{}{}{g_{WY}}) \vert$, as $39090.78$. The estimated model evidence for $m_5$ is $P(g_{WY} \vert I_{m_1} = 1) = 2.28\times10^{-27812}$. This results in posterior probabilities of $P(I_{m_3} = 1 \vert g_{WY}) \geq 99.9\%$ and $P(I_{m_5} = 1 \vert g_{WY}) \leq 0.01\%$. The posterior probability estimates for $m_3$ and $m_5$ suggest that there is strong evidence in favor of $m_3$, the model which includes only degree distribution. Again, this conclusion does not imply that degree mixing is not a significant influence on the network structure; it only means that $m_5$--as specified by our choice of $P_{\phi_5}$--did not lead to a better model compared to our choice of model for degree distribution.

\subsection{Findings Across all 50 Patient-sharing Networks}

For 47 of the 50 states, the posterior probabilities for $m_2$ when compared to $m_1$ were above $94.0\%$; the only exceptions were California ($13.2\%$), Delaware ($50.2\%$), and Wyoming ($82.6\%$). California has the lowest network density among all 50 states, whereas Delaware and Wyoming have the highest. Across all states we see strong evidence of heterogeneity in sociality. However, we see little evidence of selective mixing by provider type and degree; for computational reasons, we only investigated $10$ states for selective mixing by degree. 

\section{Discussion}

Two factors that we describe above allow investigators to make best use of their prior knowledge about networks and of observed data: 1) The ability to select the probability distribution for network properties, which enables investigators to evaluate the appropriateness of different network models that correspond to mechanisms that they hypothesize  lead to generation of an observed network, and 2) the ability to incorporate prior information into network analyses. We demonstrate these capabilities in our investigation of generative mechanisms associated with patient-sharing networks. In particular, we investigated heterogeneity in sociality as well as selective mixing by provider type and degree. To do so, we develop CCMs corresponding to these network properties. Using our Bayesian model selection approach, we found evidence in support of heterogeneity in sociality but not selective mixing. 

There are several limitations to our analysis. First, our conclusion of whether selective mixing by provider type or degree is a significant influence on the network structure was inconclusive. This result stems from the fact that there are potentially many probability distributions (beyond the ones we selected) that can be used to model mixing patterns. Therefore, more work is required to develop probability distributions for network properties that correspond to distinct generative network mechanisms; this need is particularly great for degree mixing as it is a high dimensional property, i.e., contains a large number of entries. Second, our analysis did not account for heterogeneity in the number of patients treated by providers; developing methods on how to incorporate this information into analyses is another promising area of research. Third, we investigated patient-sharing networks based on the representation of connections between providers as being binary (present or absent); future research is needed to expand CCMs to account for weighted edges.

\section*{Data Availability Statement}

The data that support the findings of this study are openly available in CMS Data portal at https://www.cms.gov/Regulations-and-Guidance/Legislation/FOIA/Referral-Data-FAQs and https://www.cms.gov/Research-Statistics-Data-and-Systems/Statistics-Trends-and-Reports/Medicare-Provider-Charge-Data/Physician-and-Other-Supplier2015.

\section*{Acknowledgments}

This research is supported by a grant from the National Institute of Health (R37 AI-51164). Conflict of Interest: None declared. 

\bibliographystyle{unsrt}  
\bibliography{paper2_bib}

\end{document}